\input{aipcheck.tex}

\documentclass[final]{aipproc}

\newcommand\doingARLO[2][]{%
  \ifx\mmref\undefined #1\else #2\fi
}

\layoutstyle{8x11double}
\begin{document}

\title {New determination of the $N$-$\Delta(1232)$ axial form factors from weak 
pion production and coherent pion production off nuclei at 
T2K and MiniBooNE energies revisited}



\author{E. Hern\'andez}{
  address={Departamento de F\'{\i}sica Fundamental e IUFFyM, Universidad de
  Salamanca, E-37008 Salamanca, Spain},
}
\author{J. Nieves}{
  address={Instituto de F\'{\i}sica Corpuscular (IFIC), Centro Mixto
  CSIC-Universidad de Valencia, Institutos de Investigaci\'on de Paterna,
  Aptd. 22085, E-46071 Valencia, Spain},
}
\author{M. Valverde}{
  address={Research Center for Nuclear Physics (RCNP), Osaka University,
Ibaraki 567-0047, Japan},
}
\author{M.J. Vicente-Vacas}{
  address={Departamento de F\'{\i}sica Te\'orica e IFIC, Centro Mixto
  CSIC-Universidad de Valencia, Institutos de Investigaci\'on de Paterna,
  Aptd. 22085, E-46071 Valencia, Spain},
}
\begin{abstract} We
 re-evaluate our model predictions in Phys.\ Rev.\ D {\bf 79}, 013002
 (2009) for different observables in neutrino induced coherent pion production.
 This comes as a result  of the new improved fit to old bubble chamber
 data  of the dominant axial  $C_5^A$ nucleon-to-Delta form factor. 
 We find an increase of 20\%$\sim$30\% in the values   for the total cross 
 sections. Uncertainties  induced by the errors
 in the determination of $C_5^A$ are computed.  Our new
 results turn out to be compatible within about $1\sigma$ with the
 former ones. We also 
 stress the existing tension between the recent experimental
 determination of the ${\sigma({\rm CC coh}\pi^+)}/{\sigma({\rm NC
 coh}\pi^0)}$ ratio by the SciBooNE Collaboration and the theoretical
 predictions.

\end{abstract}

\date{\today}

\maketitle

\section{Introduction}
Pion production by neutrinos in the
intermediate energy region is a source of relevant data on hadronic
structure. Pions are mainly produced through resonance excitation and
these reactions can be used to extract information on
nucleon-to-resonance axial transition form factors. Besides, an 
understanding of these processes is very important in the analysis of
neutrino oscillation experiments. 

 The best 
  available information on pion production by neutrinos off the nucleon
comes from old bubble chamber neutrino scattering experiments at
ANL~\cite{anlviejo,anl} and
BNL~\cite{bnlviejo,bnl}. The latter experiment provides larger cross sections
and
it has been argued in Ref.~\cite{Graczyk:2009qm} that this could be due to 
neutrino flux
uncertainties in both experiments. Both ANL and BNL data were obtained
 in deuterium\footnote{Although deuteron effects were evaluated
in Ref.~\cite{AlvarezRuso:1998hi},
where they were estimated to reduce the cross section by 5$\sim$10\%, their
effect has been neglected in most calculations to date. }, and at relatively 
low energies for which the
dominant contribution is given by the $\Delta$ pole ($\Delta P$)
mechanism: weak excitation of the $\Delta(1232)$ resonance and its
subsequent decay into $N\pi$. 
A convenient parameterization of the $W^+n\to\Delta^+$ vertex is given
in terms of eight  form-factors: four vector and four
axial ($C^A_{3,4,5,6}$ ) ones.  Vector form factors have been
determined from the analysis of photo and electro-production
data and in our work we use the parameterization of Lalakulich {\it et
  al.}~\cite{Lalakulich:2006sw}.  Among the axial form factors the most
important contribution comes from $C_5^A$. The form factor $C_6^A$,
which contribution to the differential cross section vanishes for
massless leptons, can be related to $C_5^A$ thanks to the partial
conservation of the axial current. Since there are no other theoretical
constraints for $C_{3,4,5}^A(q^2)$, they have to be fitted to data.
Most analyses, including the ANL and BNL ones, adopt Adler's
model~\cite{adler} where $C_3^A(q^2) = 0$ and $C_4^A(q^2)
= -{C_5^A(q^2)}/{4}$.
\begin{figure}[htb]
\includegraphics[scale=0.2]{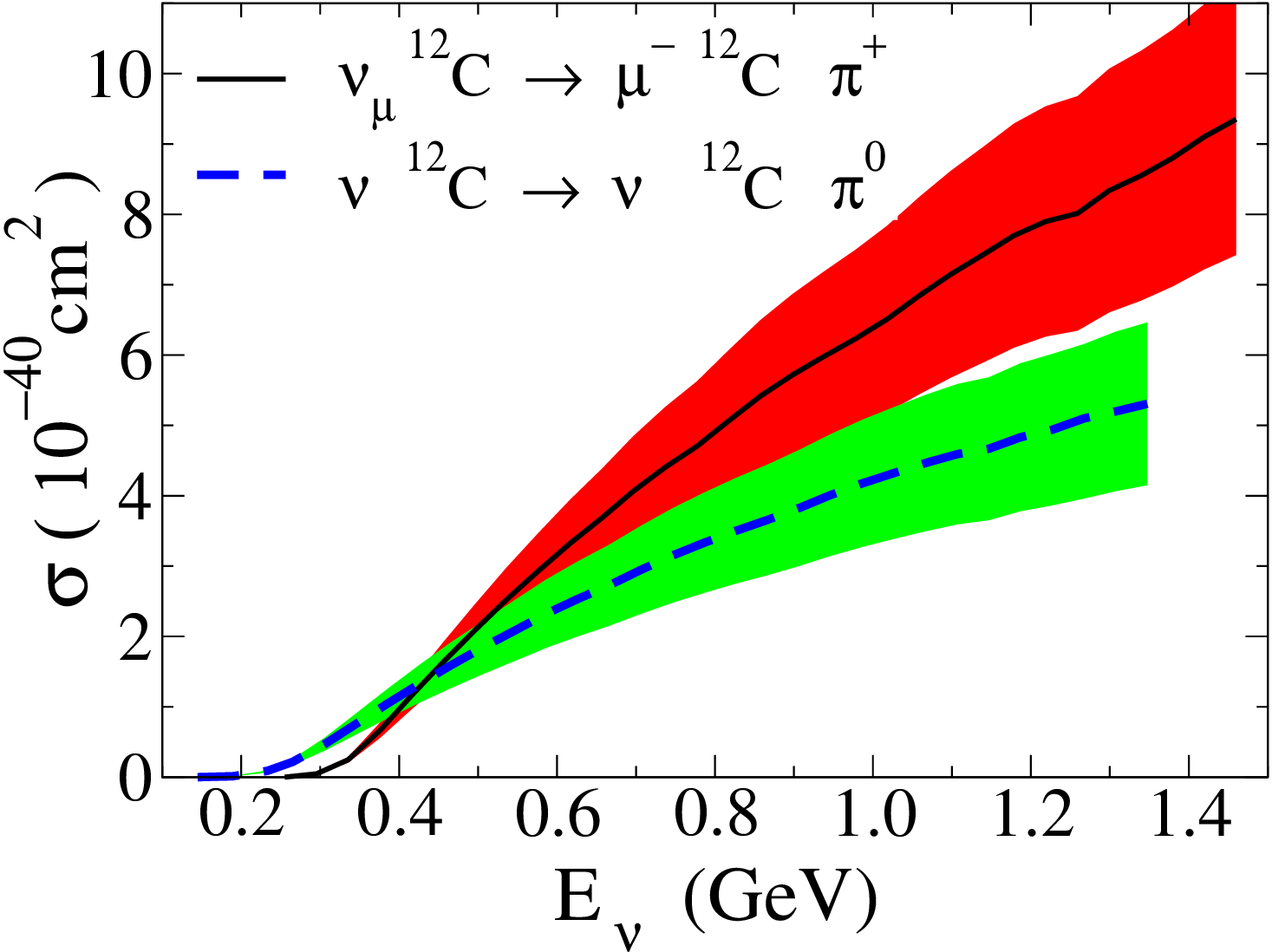}\hspace{.5cm}
\includegraphics[scale=0.2]{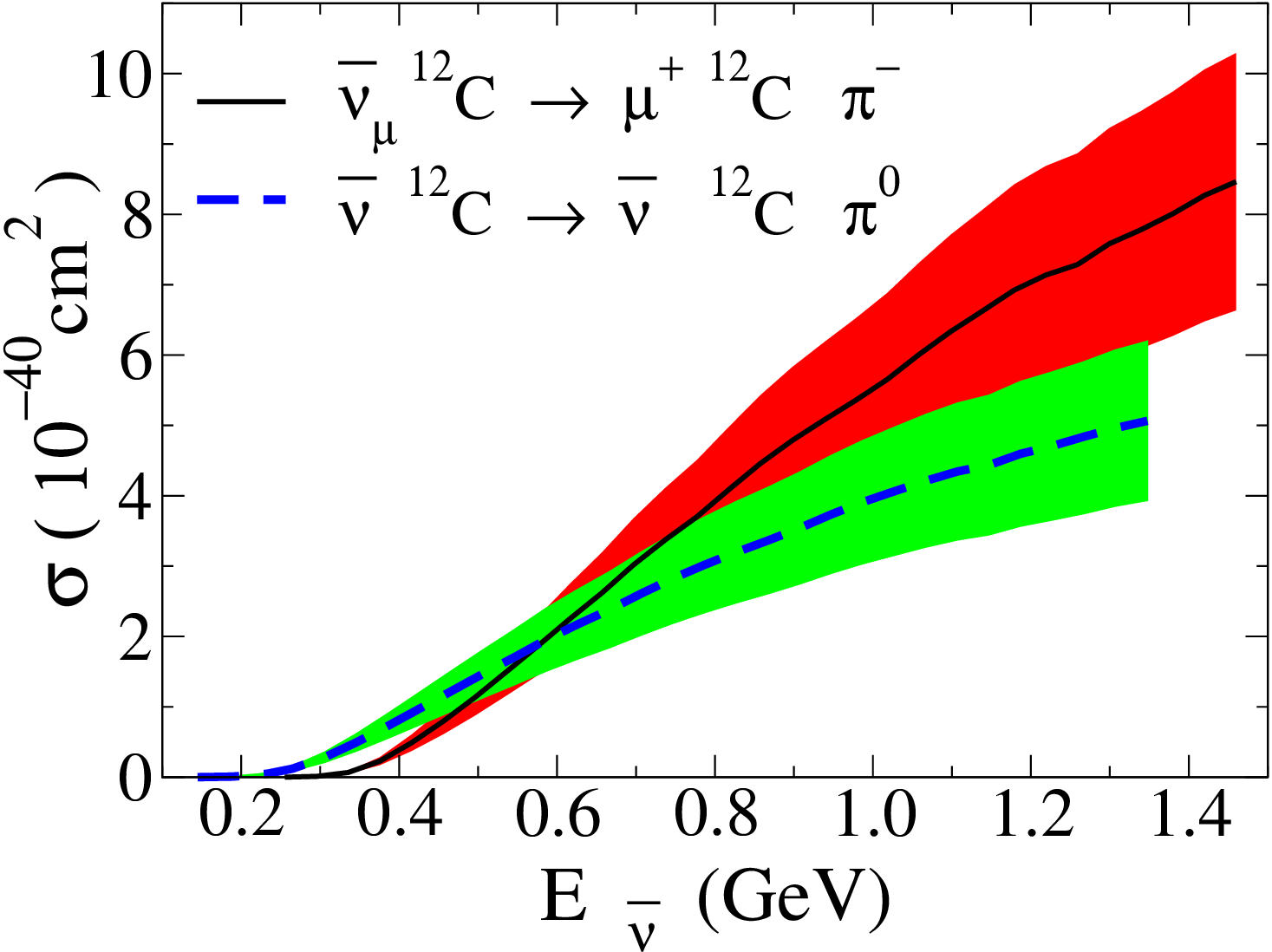}
\hspace{.1cm}\includegraphics[scale=0.2]{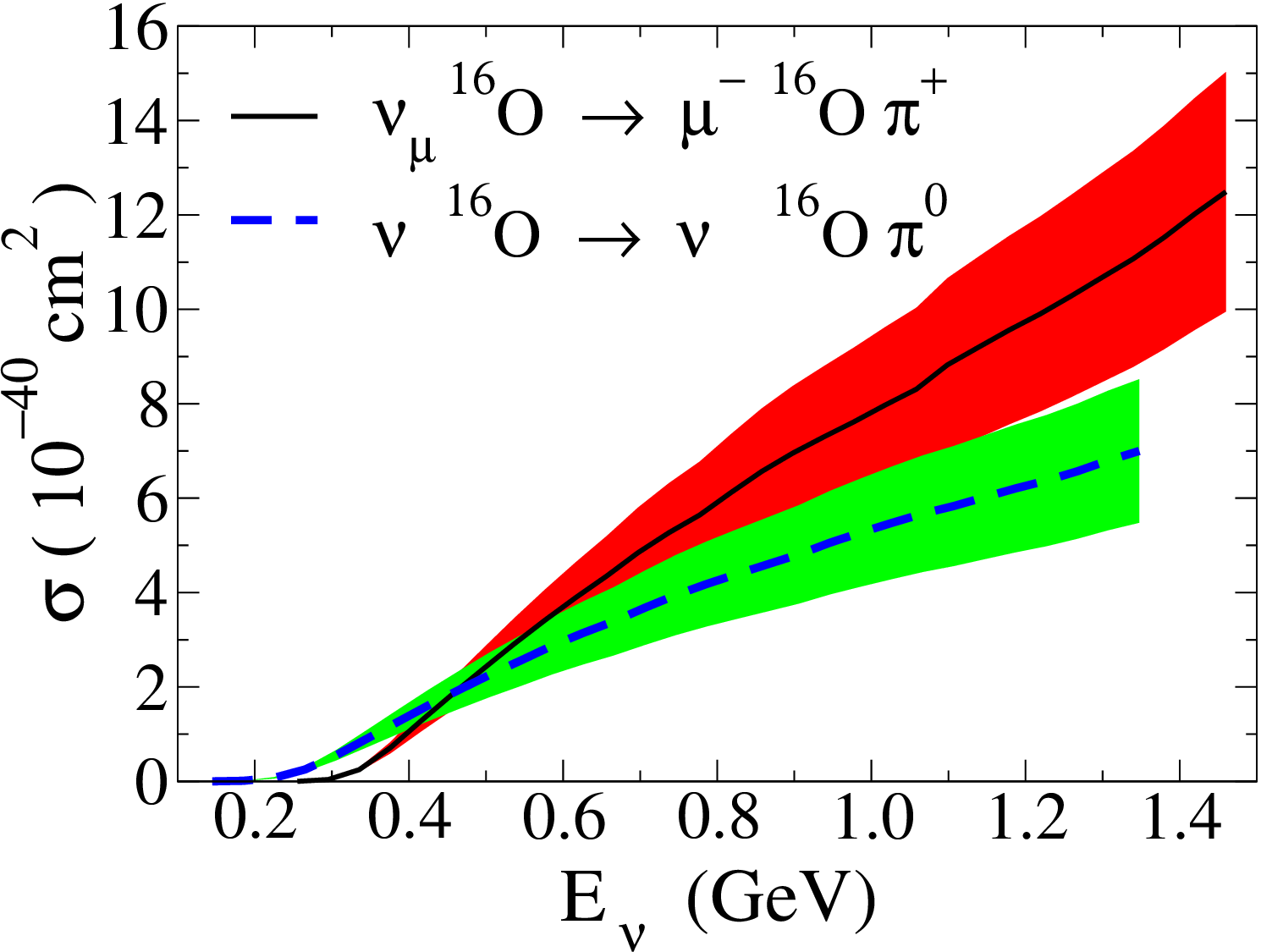}\\ 
\caption{ $\nu_\mu/\bar\nu_\mu$ CC and $\nu/\bar\nu$  NC\ C$\pi$P cross sections
 from a carbon target
 (left \& middle panels) and $\nu_\mu$
 CC and $\nu$ NC\  C$\pi$P cross sections from an oxygen target (right panel)
as a function of the neutrino/antineutrino energy.  Error bands are
shown. Taken from Ref.~\cite{Hernandez:2010jf}.}\label{fig:cross}
\end{figure}

In Ref.\cite{Hernandez:2007qq} we developed a model for  weak pion production
 off the nucleon that,
   besides the dominant $\Delta P$ mechanism,
  included   background terms required by chiral symmetry
  \footnote{Some background terms were
already considered in the works of Ref.~\cite{backg}.}. Background terms 
produce quite significant effects and as a result  we had
to re-fit the $C_5^A(q^2)$ form factor using for that purpose the flux averaged $\nu_\mu p
\to \mu^-p\pi^+$ ANL $q^2-$differential cross section data with final
pion-nucleon invariant mass $W<
1.4\,$GeV. We found $C_5^A(0) = 0.867 \pm 0.075$\ \footnote{Results for the
$M_{A\Delta}$ axial mass for this and other fits discussed in this contribution
can be found in the  quoted references.}. This represents a  correction 
of the order of
30\% to the off diagonal Goldberger-Treiman  relation (GTR)
prediction of $C_5^A(0) \approx 1.2$,  the latter being the values  used in most
theoretical analyses. The consideration of background terms, with the reduced
$C_5^A(0)$ value,
gave rise to
 an overall improved
description of the data, as compared to the case where only the
$\Delta P$ mechanism was considered. As mentioned, there is some degree 
 of inconsistency among
the ANL and BNL measurements and  a good description of BNL data 
requires a  $C_5^A(0)$ value close to the GTR one. The model, with the
corresponding medium modifications, was latter applied 
to coherent pion production (C$\pi$P) in nuclei in Ref.~\cite{Amaro:2008hd}. 
C$\pi$P is a  low $q^2$ process which is dominated by the 
axial part of the weak current.  Besides, background term
contributions  cancel to a large extent for symmetric
nuclei, being the $\Delta P$  mechanism the
unique contribution. C$\pi$P  is thus  very sensitive to 
$C_5^A(0)$ and our predictions for C$\pi$P cross sections 
in Ref.~\cite{Amaro:2008hd} were notably smaller
than  previous existing calculations.

Last year  Graczyk {\it et al.}~\cite{Graczyk:2009qm} made a combined
fit of both ANL and BNL data taking into account 
 deuteron effects,  evaluated as in Ref.~\cite{AlvarezRuso:1998hi}, 
and, most important, the uncertainties in the neutrino flux
normalization, the idea being that ANL and BNL are not incompatible when flux
uncertainties are taken into account. 
Flux uncertainties  were treated as systematic errors
 and taken to be 20\% for ANL
data and 10\% for BNL data. With a pure dipole dependence for $C_5^A$, 
they found $C_5^A(0)=1.19 \pm 0.08$, in agreement with
the GTR estimate. 
The work in Ref.~\cite{Graczyk:2009qm} considered only 
the $\Delta P$  mechanism
but ignored the sizable non-resonant (background) contributions
which are of special relevance for  neutrino energies below 1 GeV.

In Ref.~\cite{Hernandez:2010bx}, we have conducted a similar analysis,
that besides including ANL and BNL data,  deuteron effects and flux 
uncertainties,   it also
takes into account the effects of background terms. We now
get $C_5^A(0)=1.00 \pm 0.11$, which is just 2$\sigma$ away from the GTR value.
As a consequence of this new value we  expect  our  results 
 in
Ref.~\cite{Amaro:2008hd}  to underestimate C$\pi$P cross sections by some
30\% (corresponding to
 $[\left.C_5^A(0)\right|_{new}/\left.
C_5^A(0)\right|_{former}]^2=[1/0.867]^2=1.33$). In this  contribution we show
some new results, with theoretical uncertainties, for C$\pi$P 
observables obtained with the new
 $C_5^A(0)$ determination. Further results  can be found in 
 Ref.~\cite{Hernandez:2010jf}.

In Fig.~\ref{fig:cross} we show new $\nu_\mu/\bar\nu_\mu$ charged current (CC)
 and $\nu/\bar\nu$  neutral current (NC) C$\pi$P 
cross sections from a carbon target
 (left \& middle panels) and $\nu_\mu$
 CC and $\nu$ NC C$\pi$P cross sections from an oxygen target. 
 We find an increase 
in the cross sections  by some 20-30\%, depending on energy,  with respect to our former
results in Ref.~\cite{Amaro:2008hd}. This is a consequence 
 of the new $C_5^A(0)$ value . As in Ref.~\cite{Amaro:2008hd},
there are 
large deviations from the approximate relation $\sigma_{\rm CC} \approx 2 
\sigma_{\rm NC}$ for
these two isoscalar nuclei in the whole range of $\nu/\bar\nu$
energies examined.
This is greatly due to the
finite muon mass~\cite{rein2,Berger:2007rq}, and thus the deviations 
are dramatic at low neutrino
energies. 

\begin{table}
    \begin{tabular}{lcccccc}\hline
Reaction                 & Experiment &\hspace*{1cm}$\bar \sigma$\hspace*{1cm}&\ \ 
$\sigma_{\rm exp}$\ \ &
$E_{\rm max}^i $ ~~ & $\int_{E_{\rm low}^i}^{E_{\rm max}^i} dE \phi^i(E)
\sigma(E)$~~ & $\int_{E_{\rm low}^i}^{E_{\rm max}^i} dE \phi^i(E)$
\\
&  & [$10^{-40}$cm$^2$] & [$10^{-40}$cm$^2$] & [GeV]
&[$10^{-40}$cm$^2$]  \\\hline
CC\phantom{*} $\nu_\mu + ^{12}$C    & K2K        &   $6.1\pm1.3$    &
$<7.7 $~\cite{Hasegawa:2005td}   &1.80&$5.0\pm1.0$&0.82         \\
CC\phantom{*} $\nu_\mu + ^{12}$C    & MiniBooNE  &   $3.8\pm0.8$    
&          &1.45&$3.5\pm0.7$&0.93 \\

CC\phantom{*} $\nu_\mu + ^{12}$C    & T2K        &   $3.2\pm0.6$   
 &          &1.45&$2.9\pm0.6$&0.91           \\
CC\phantom{*} $\nu_\mu + ^{16}$O    & T2K        &   $3.8\pm0.8$    &    &
  1.45& $3.4\pm0.7$&  0.91           \\
NC\phantom{*} $\nu_{{\mu}} + ^{12}$C    & MiniBooNE  & $2.6\pm0.5$ &
 $7.7\pm1.6\pm3.6$~\cite{Raaf}
&1.34&$2.2\pm0.5$&0.89\\
NC\phantom{*} $\nu_{{\mu}} + ^{12}$C    & T2K        & $2.3\pm0.5$    
 &     &1.34 &$2.1\pm0.5$
&0.90 \\
NC\phantom{*} $\nu_{{\mu}} + ^{16}$O    & T2K        & 
  $2.9\pm0.6$      & &  1.35     &  $2.6\pm0.6$ &  0.90  \\
CC\phantom{*} $\bar\nu_\mu + ^{12}$C    & T2K        &   $2.6\pm0.6$ &
&    1.45      & $1.8\pm0.4$ & 0.67          \\
NC\phantom{*} $\bar\nu_{{\mu}} + ^{12}$C    & T2K        &
 $2.0\pm0.4$     &     &1.34 &$1.3\pm0.3$
&0.64 \\
\hline
    \end{tabular}
  \caption{\footnotesize NC/CC $\nu_\mu$ and $\bar\nu_\mu$ coherent
    pion production total cross sections, with errors, for K2K,
    MiniBooNE and T2K experiments. In the case of CC K2K, the
    experimental threshold for the muon momentum $|\vec{k}_\mu|>$
    450 MeV is taken into account.    Details on the flux convolution are compiled in
    the last three columns. Taken from Ref.~\cite{Hernandez:2010jf}.}
\label{tab:res} 
\end{table}

In Table~\ref{tab:res} we show  new predictions for both NC 
and CC
processes, for the
K2K~\cite{Hasegawa:2005td} and MiniBooNE~\cite{AguilarArevalo:2008xs}
flux averaged cross sections as well as for the  T2K
experiment.   As in Ref.~\cite{Amaro:2008hd}, and
since we neglect all resonances above the $\Delta(1232)$, we have set
up a maximum neutrino energy ($E_{\rm max}^i$) in the flux
convolution, of $E_{\rm max}=1.45\,$GeV and 1.34 GeV for CC
and NC $\nu_\mu/\bar\nu_\mu$ driven processes, respectively.
For the K2K case a threshold of 450 MeV for muon
momentum is also implemented~\cite{Hasegawa:2005td}  and  we 
can to go up to $E_{\rm max}^{\rm
CC,K2K}$=1.8 GeV. In this way we  cover about 90\% of
the total flux in most of the cases. For the T2K antineutrino flux, we
cover just about 65\%, and therefore our results
 are less reliable.
Central value cross sections increase by some 23\%$\sim$30\%, while the
errors, associated to the uncertainties in the $C_5^A(q^2)$ 
 determination, are of the order of 21\%. Our new results
are thus compatible with former ones in Ref.~\cite{Amaro:2008hd}
within $1\sigma$. 
 Our prediction for the K2K experiment lies more
than 1$\sigma$ below the K2K upper bound, while we still predict an NC
MiniBooNE cross section notably smaller than that given in the PhD
thesis of J.L. Raaf~\cite{Raaf}. Note however, that  this not an official number
by the MiniBooNE Collaboration.

Finally we address the issue of the 
$\frac{\sigma({\rm CC coh}\pi^+)}{\sigma({\rm NC coh}\pi^0)}$ ratio.
The SciBooNE Collaboration has just reported a measurement of NC
coherent $\pi^0$ production on carbon by a $\nu_\mu$ beam with average
energy 0.8\,GeV~\cite{sciboone}. Based on previous measurements of CC
coherent $\pi^+$ production~\cite{sciboone2}, they conclude that
%
$\left.\frac{\sigma({\rm CC coh}\pi^+)}{\sigma({\rm NC coh}\pi^0)}\right|_{\rm
SciBooNE}=0.14^{+0.30}_{-0.28}$.
%
This result can not be accommodated within our model, or any other
present theoretical model. Isospin symmetry would predict an exact value of 2 for
this ratio in  isoscalar
nuclei,   like carbon or oxygen, provided that
1) vector current is neglected, 2) the muon mass is neglected, and 3)
 $\cos\theta_C= 1$. The vector current contribution is known to be suppressed 
 by the nuclear form
factor (see the discussion after Eq.(5) in Ref.~\cite{Amaro:2008hd}) 
and the effect of the muon mass can not explain  the SciBooNE result for
an average  neutrino energy of $0.8\,$GeV.
Besides, $\cos\theta_C= 0.974$. Theoretically, one would not expect
this ratio to be
  much smaller than 1.4-1.6. For instance, for a carbon target and for
a neutrino energy of 0.8\,GeV we find a value of $1.45\pm 0.03$ for that ratio,
ten times bigger that the value given by the SciBooNE Collaboration.
 From the $\nu_\mu+^{12}$C CC and NC MiniBooNE convoluted results shown
in Table~\ref{tab:res} we obtain $1.46 \pm 0.03$.  
We believe part of this huge
discrepancy with the SciBooNE result stems form the use in Ref.~\cite{sciboone} 
of the Rein-Sehgal~\cite{rein,rein2} model
to estimate the ratio between NC coherent $\pi^0$ production and the
total CC pion production. As clearly shown in
Refs.~\cite{Amaro:2008hd,Hernandez:2009vm}, the RS model is not
appropriate to describe coherent pion production in the low energy
regime of interest for the SciBooNE experiment.

\begin{theacknowledgments}
M.V. acknowledges
support from the Japanese Society for the Promotion of Science.  Work supported by DGI and FEDER funds, contracts
 No. FIS2008-01143/FIS, No. FIS2006-03438, No. FPA2007-65748, No. CSD2007-00042, by JCyL,  
  contracts No. SA016A07 and No. GR12,  by GV, contract No. PROMETEO/2009-0090 and by the
   EU   HadronPhysics2 project, contract No. 227431. 

\end{theacknowledgments}


\doingARLO[\bibliographystyle{aipproc}]
          {\ifthenelse{\equal{\AIPcitestyleselect}{num}}
             {\bibliographystyle{arlonum}}
             {\bibliographystyle{arlobib}}
          }
\bibliography{sample}

\end{document}